\documentclass[conference]{IEEEtran}
\IEEEoverridecommandlockouts
\usepackage{amsmath,amssymb,amsfonts}
\usepackage{algorithmic}
\usepackage{graphicx}
\usepackage{textcomp}
\usepackage[table,xcdraw]{xcolor}
\usepackage{balance}
\usepackage{soul}
\usepackage{booktabs}
\usepackage{multirow}
\usepackage{float}
\usepackage{url}
\usepackage{tcolorbox}
\usepackage{array}
\usepackage{tabularx} 
\usepackage[noadjust]{cite}
\usepackage[bookmarks=true,pagebackref=true,hidelinks]{hyperref}
\usepackage{xurl}
\usepackage{enumitem}
\setlist[itemize,enumerate]{noitemsep, topsep=0pt, leftmargin=1.5em}
\usepackage[utf8]{inputenc}
\usepackage{newunicodechar,graphicx}
\DeclareRobustCommand{\okina}{%
  \raisebox{\dimexpr\fontcharht\font`A-\height}{%
    \scalebox{0.8}{`}%
  }%
}
\newunicodechar{ʻ}{\okina}
\usepackage{background}
\backgroundsetup{
  position=current page.east,
  angle=-90,
  nodeanchor=east,
  vshift=-3mm,
  opacity=1,
  scale=3,
  contents=PREPRINT
}

\newcommand{\RQA}{\textbf{RQ1}: How effectively can LLMs identify and classify grammatical patterns in method names?}

\newcommand{\RQB}{\textbf{RQ2}: To what extent are LLM-suggested method name corrections aligned with software engineering best practices?}

\begin{document}

\title{Exploring Large Language Models for Analyzing and Improving Method Names in Scientific Code}

\author{
    \IEEEauthorblockN{Gunnar Larsen}
    \IEEEauthorblockA{\textit{University of Hawaiʻi at Mānoa} \\
    Hawaiʻi, USA \\
    gunnarrl@hawaii.edu}
    
    \and
    \IEEEauthorblockN{Carol Wong}
    \IEEEauthorblockA{\textit{University of Hawaiʻi at Mānoa} \\
    Hawaiʻi, USA \\
    carolw8@hawaii.edu}
    
    \and
    \IEEEauthorblockN{Anthony Peruma}
    \IEEEauthorblockA{\textit{University of Hawaiʻi at Mānoa} \\
    Hawaiʻi, USA \\
    peruma@hawaii.edu}
}

\maketitle

\begin{abstract}
Research scientists increasingly rely on implementing software to support their research. While previous research has examined the impact of identifier names on program comprehension in traditional programming environments, limited work has explored this area in scientific software, especially regarding the quality of method names in the code. The recent advances in Large Language Models (LLMs) present new opportunities for automating code analysis tasks, such as identifier name appraisals and recommendations. Our study evaluates four popular LLMs on their ability to analyze grammatical patterns and suggest improvements for 496 method names extracted from Python-based Jupyter Notebooks. Our findings show that the LLMs are somewhat effective in analyzing these method names and generally follow good naming practices, like starting method names with verbs. However, their inconsistent handling of domain-specific terminology and only moderate agreement with human annotations indicate that automated suggestions require human evaluation. This work provides foundational insights for improving the quality of scientific code through AI automation.

\end{abstract}


\section{Introduction}
\label{Section:intro}
As Large Language Models (LLMs) grow in popularity, their application in software engineering is becoming increasingly significant \cite{Ebert2023}. These models assist developers with various tasks, such as code generation, debugging, code review assistance, and documentation \cite{Hou2024,Liang2024,Sauvola2024,Belzner2023}. Additionally, through the use of IDE plugins like GitHub Copilot\footnote{\url{https://github.com/features/copilot}} and specialized AI-integrated IDEs such as Cursor\footnote{\url{https://cursor.com/}} and Replit\footnote{\url{https://replit.com/}}, developers can easily leverage LLMs to work faster and more efficiently.

Despite these advancements, there are also drawbacks associated with using AI-powered tools in software engineering \cite{Gao2025}. Code quality issues, including security vulnerabilities, potential biases in the generated code, and hallucinations, lead to developers losing trust in AI-generated code, as they often need to double-check and correct the output \cite{Mozannar2024, Lertbanjongngam2022}. Additionally, crafting prompts can be time-consuming, and latency or interruptions may disrupt workflow. Moreover, there are also concerns regarding the ethical and legal implications of AI-generated content \cite{Bull2024,Russo2024,Fan2023}.

While software engineers have the necessary education and experience to effectively utilize these LLM tools and evaluate their output, individuals without a software engineering background may inadvertently use code generated by LLMs without fully understanding how it could affect the overall quality of the codebase and future maintenance efforts. This includes research scientists, who often need to develop software programs to support their research. 
Typically, scientists working in non-computer science research domains learn programming through on-the-job training or from informal resources like online articles and blogs, which can lead to a lack of understanding regarding the principles of writing readable and maintainable code \cite{Chen2025}. 

Prior research into the code typically written by research scientists reveals issues such as PEP8 violations, stylistic inconsistencies, high coupling, and significant challenges in reproducibility \cite{Wang2020,grotov2022large,Adams2023,PimentelMSR2019}. Furthermore, Wong et al. \cite{Wong2025} examined the method names in such code and found many instances where names deviated from standard naming practices, such as starting with a verb. Instead, these names tend to be based on the output of the methods, rather than their actions. The authors also noted the presence of ambiguous single-term names, unconventional word order, and methods that include abbreviations or acronyms, many of which require domain-specific knowledge for interpretation. They also observed instances where the names incorrectly ended with a verb.

\subsection{Goal and Research Questions}
In this context, our exploratory study aims to investigate the ability of LLMs to evaluate method names in scientific code and to generate alternative naming suggestions when appropriate. The naming conventions observed in the code written by research scientists frequently diverge from standard practices, leading to reduced clarity and maintainability. Method names, in particular, play an important part in conveying the purpose and functionality of code, and poorly chosen names can result in confusion and hinder collaboration among researchers \cite{Hst2009,Host2007,Alsuhaibani2021}. By utilizing the capabilities of LLMs, we seek to determine how effectively these models can understand context and propose method names that adhere to established conventions and best practices. We envision our findings advancing the knowledge of identifier naming in scientific code and the effectiveness and drawbacks of LLMs in generating improved naming suggestions. Additionally, it offers practical guidelines for non-computer scientists when using LLMs for code generation and evaluation. We address the following research questions (RQs):

\vspace{1.3mm}
\noindent\textbf{\RQA} Method names in scientific code typically do not follow standard software engineering naming conventions. This RQ aims to explore the effectiveness of LLMs in generating part-of-speech tags for these method names. This initial understanding will help us determine the reliability of LLMs as tools for analyzing the grammatical structure of method names in scientific code.

\vspace{1.3mm}
\noindent\textbf{\RQB} In this RQ, we aim to evaluate the quality and appropriateness of method name improvements suggested by LLMs for scientific code, including how they handle abbreviations and acronyms contained in the name. Insights gained from this exploration will assist us in developing better tools and techniques to improve code quality in notebook environments.

\subsection{Contributions}
The main contributions of this work are:
\begin{itemize}
    \item Evaluation of LLMs: Assesses the ability of LLMs to analyze grammatical patterns in method names within scientific programs and suggest corrections aligned with software engineering best practices.
   \item Advancing Code Quality: Contributes to improving the quality of scientific programs, thereby enhancing their maintainability and reproducibility.
\end{itemize}

\section{Methodology}
\label{Section:method}
\begin{figure*}
 	\centering
 	\includegraphics[trim=0cm 0.2cm 0cm 0cm,clip,scale=0.7]{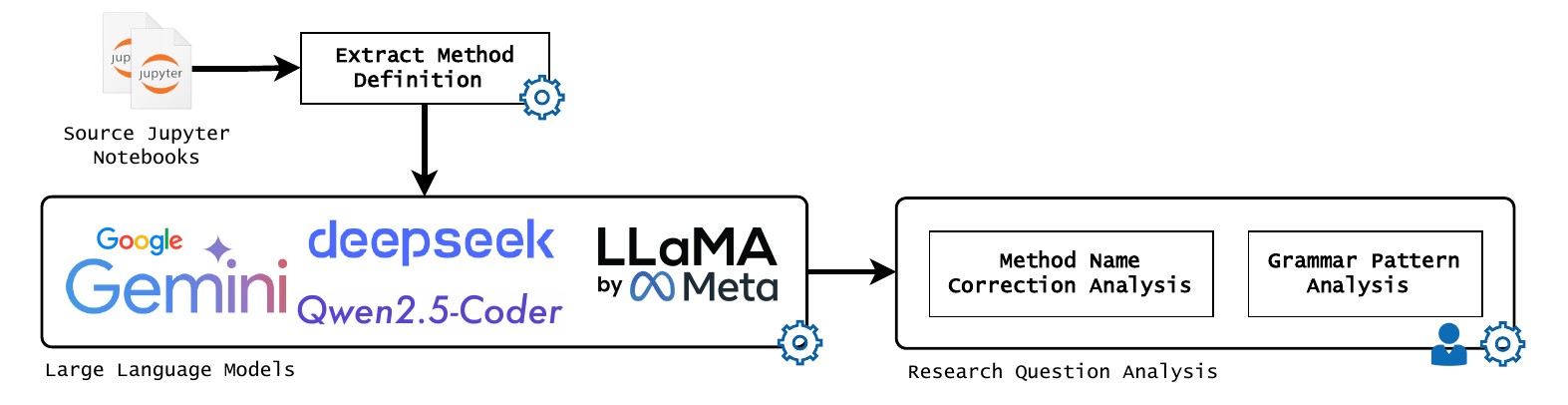}
    \caption{KEY ACTIVITIES IN OUR METHODOLOGY.}
    \vspace{-3mm}
    \label{Figure:method}
\end{figure*}

In this section, we describe in detail our approach to answering our RQs. Figure \ref{Figure:method} depicts an overview of the key activities in our methodology, which we elaborate below.

\subsection{Source Dataset}
In this study, we utilize the dataset made available by Wong et al. \cite{Wong2025}. In their study, the authors manually reviewed 691 method names extracted from 384 Jupyter Notebooks, which were randomly sampled from a larger dataset of 847,881 Python-based Jupyter Notebooks, which were collected from public GitHub repositories by Grotov et al. 
\cite{grotov2022large}. The authors (i.e., Wong et al.) annotated these method names for grammatical patterns, using part-of-speech tags, defined by Newman et al. \cite{NewmanCoRR2020}, specifically tailored for analyzing source code identifiers. This dataset  
is the only one that has part-of-speech annotated method names from scientific code. 

\subsection{Part-of-Speech Tags}
These part-of-speech tags utilized by Wong et al. include nouns (N), noun modifiers (NM) for adjectives and noun-adjuncts, verbs (V), verb modifiers (VM), prepositions (P), determiners (DT), conjunctions (CJ), pronouns (PR), digits (D), and preambles (PRE). Details about the tags are available in the study by  Newman et al. \cite{NewmanCoRR2020} and online at \cite{catalog}.

\subsection{Method Extraction}
For each of the 691 method names in the source dataset, we extracted the complete method definition from its respective notebook file. We utilized the \texttt{nbformat} Python package to parse the notebook and extract the method code. 

\subsection{Large Language Model Analysis}
For this study, we utilized four popular Large Language Models (LLMs). Each of these LLMs  
has been shown by its respective authors to demonstrate significant capabilities in understanding and generating code-related tasks. The models we selected are:
\begin{itemize}
    \item Google Gemini 2.0 Flash\footnote{\url{https://ai.google.dev/gemini-api/docs/models\#gemini-2.0-flash}}
    \item Alibaba Qwen 2.5 Coder 32B\footnote{\url{https://ollama.com/library/qwen2.5-coder}}
    \item Meta LLaMa 3.3 70B\footnote{\url{https://ollama.com/library/llama3.3}}
    \item DeepSeek-R1 70B\footnote{\url{https://ollama.com/library/deepseek-r1}}
\end{itemize}

We built custom scripts to interact with the Gemini model using Google's Gemini API. For the remaining three models, we utilized the Ollama\footnote{\url{https://ollama.com/}} platform to run the LLMs locally. 

\subsection{Prompt Design}
The effectiveness of LLM-based analysis depends on carefully crafted prompts, as their phrasing, structure, and context influence output quality and consistency \cite{sahoo2024systematic}. Hence, we evaluated multiple variations of the prompt to determine the most effective. Our evaluation included prompts with different roles, levels of detail, various examples, and alternative formatting approaches. We found that prompts with clear structural guidelines and explicit output formatting requirements produced the most consistent results across different LLM architectures. Our final prompt design includes the following elements:
\begin{itemize}
    \item A role specification for the LLM as ``an expert software engineer specializing in Python programming''
    \item A detailed explanation of the part-of-speech tagging process, including a comprehensive table of tags (N, V, NM, VM, P, etc.) with examples
    \item Instructions for splitting method names into individual terms based on camelCase, PascalCase, or snake\_case 
    \item Guidelines for handling acronyms, abbreviations, and numerical elements in method names
    \item Explicit directions for evaluating the quality of method names against software engineering best practices
    \item A structured JSON response format with fields for:
    \begin{itemize}
        \item Current method name and its grammar pattern
        \item Corrected method name (if needed)
        \item Corrected grammar pattern (if needed)
    \end{itemize}
    \item An example of the output structure
    \item The code associated with the method 
\end{itemize}

We observed the need to provide explicit instructions regarding the part-of-speech tagging, as without such explicit guidance, the LLMs default to standard natural language part-of-speech tags, which are inadequate for analyzing identifier names in code. This approach ensures that the LLMs use a consistent tagging scheme and facilitates ease of comparison against the human-annotated tags in the source dataset.   

\subsection{Research Question Analysis}
The output generated from the LLMs was saved to a SQLite database in a table whose fields correspond to the output structure specified in the prompt.

For our analysis, we employed standard statistical measures to effectively summarize the data. We also wrote SQL queries and custom code to retrieve information from the database to address our RQs. Additionally, we manually searched the database to include representative examples that would enhance our analysis. To evaluate the level of agreement among the LLMs and between the original annotations, we utilized Fleiss' Kappa and Cohen's Kappa.  Both of these measures assess inter-rater reliability, but they differ in their application: Cohen's Kappa is appropriate for situations involving two raters, while Fleiss' Kappa is suitable for three or more raters.

\noindent Artifacts for this study are available at \cite{website}.

\section{Results}
\label{Section:results}
\begin{table}
\centering
\caption{Accuracy and agreement rates between each LLM and the human annotation (ground truth)}
\vspace{-3mm}
\label{Table:current-grammar-human}
\resizebox{\columnwidth}{!}{%
\begin{tabular}{@{}lrr@{}}
\toprule
\multicolumn{1}{c}{Model} & \multicolumn{1}{c}{Accuracy} & \multicolumn{1}{c}{Cohen's Kappa} \\ \midrule
Gemini 2.0 Flash   & 0.611 & 0.577 \\
Qwen 2.5 Coder 32B & 0.454 & 0.412 \\
DeepSeek-R1 70B    & 0.393 & 0.344 \\
LLaMa 3.3 70B      & 0.357 & 0.295 \\ \bottomrule
\end{tabular}%
}
\end{table}

\begin{table}
\centering
\caption{The top three most common grammar pattern misclassifications for each LLM.}
\vspace{-3mm}
\label{Table:grammar-pattern-misclassifications}
\resizebox{\columnwidth}{!}{%
\begin{tabular}{@{}llll@{}}
\toprule
\multicolumn{1}{c}{\textbf{\begin{tabular}[c]{@{}c@{}}Human Annoated\\ Grammar Pattern\end{tabular}}} &
  \multicolumn{1}{c}{\textbf{\begin{tabular}[c]{@{}c@{}}Model Generated\\ Grammar Pattern\end{tabular}}} &
  \multicolumn{1}{c}{\textbf{Count}} &
  \multicolumn{1}{c}{\textbf{Percentage}} \\ \midrule
\multicolumn{4}{c}{Model: \textit{Gemini 2.0 Flash}} \\
N            & PRE        & 16       & 8.29\%        \\
VM           & V          & 11       & 5.70\%        \\
V,NPL        & V,N        & 9        & 4.66\%        \\ \midrule
\multicolumn{4}{c}{Model: \textit{Qwen 2.5 Coder 32B}}      \\
V,N          & V,NM       & 42       & 15.50\%       \\
V,NPL        & V,NM       & 15       & 5.54\%        \\
V            & N          & 13       & 4.80\%        \\ \midrule
\multicolumn{4}{c}{Model: \textit{DeepSeek-R1 70B}}         \\
V,NPL        & V,N        & 18       & 5.64\%        \\
N            & V          & 13       & 4.08\%        \\
NM,N         & N,N        & 13       & 4.08\%        \\ \midrule
\multicolumn{4}{c}{Model: \textit{LLaMa 3.3 70B}}           \\
N            & V          & 15       & 4.98\%        \\
V,N          & V          & 14       & 4.65\%        \\
V,N          & V,NM       & 12       & 3.99\%        \\ \bottomrule
\end{tabular}%
}
\end{table}

In this section, we answer our RQs by analyzing the output of the LLMs.  It is important to note that although the prompts used for the LLMs were identical, not all models produced the desired output. In some cases, certain LLMs either did not adhere to the instructions and instead returned lengthy paragraphs or followed the required output format but failed to provide values for all fields in the JSON structure. Additionally, there were also instances of the model hallucinating, generating output for methods that did not exist. Out of a total of 691 input methods, the Gemini model successfully returned complete outputs for 690 of them. In comparison, the LLaMA model produced details for 615 methods, the Qwen model for 595 methods, and DeepSeek for 526 methods. However, as mentioned above, some outputs from these models are hallucinations. After examining the outputs from each of the LLMs, we identified a total of 496 methods that contained valid outputs common to all four models. 
This number reflects our decision to exclude methods that were only returned by some models, ensuring consistency in the results we analyzed. Henceforth, our analysis is based only on these 496 methods.

\subsection*{\RQA}
This RQ examines the part-of-speech tags generated by the LLMs for the method names and compares these tags against the human-annotated tags in the original dataset. First, when assessing the level of agreement among the LLMs, we find that all four models agree in 18.8\% of cases, resulting in a Fleiss' Kappa of 0.333. This indicates a fair level of agreement among the models. However, comparing the output solely across LLMs serves only to shed light on AI behavior, not to validate correctness. Therefore, we also compared the output of each LLM against the human-annotated tags in the original dataset. In Table \ref{Table:current-grammar-human}, we show the accuracy and agreement of the part-of-speech tags for each of the four LLMs. Even though Gemini had the highest accuracy, 61.1\%, the Cohen's Kappa of 0.577 is indicative of a moderate agreement.

Our next analysis focuses on the common grammar patterns that each LLM misclassified. Table \ref{Table:grammar-pattern-misclassifications} shows the top three most common grammar patterns each LLM misclassified. From this, we observe that Gemini struggles with single-term names that are acronyms. The N $\rightarrow$ PRE misclassification affecting methods having names like ``SVD'' and ``MSE'' reveals that Gemini incorrectly categorizes mathematical and technical abbreviations as preambles. 

Qwen tends to misclassify nouns as noun modifiers (V,N$\rightarrow$V,NM and V,NPL$\rightarrow$V,NM). Unlike a noun, a noun modifier adds specificity to the noun by providing more detail and clarifying the particular type of object to which the action relates. Hence, it is interesting that even though method names like ``load\_image'' are composed of only a verb and a noun, Qwen often classifies the noun as a noun modifier. This misclassification occurs despite the fact that in this context, ``image''  acts as a direct object of ``load,'' not a modifier.

The analysis also reveals challenges with words that can function as both nouns and verbs. Qwen classifies single-word method names ``clean'' and  ``animate'' as nouns, while the human-annotated dataset has them listed as nouns. In contrast, Deepseek and LLaMA incorrectly classify certain human-annotated nouns, such as ``sigmoid'' and ``answer'', as verbs. These dual-function words require understanding of both method naming conventions and contextual behavior, not just linguistic rules, for accurate classification.

An interesting observation regarding certain models is how they classify plural words. For example, human annotations for method names like ``build\_data'' and ``process\_features'' categorize them as V,NPL, meaning a verb followed by a plural noun. However, models such as Gemini, Qwen, and Deepseek classify the plural term as singular (i.e., V,N).

\vspace{1mm}
\noindent\textbf{Summary.}
RQ1 examines how effectively LLMs classify grammatical patterns in method names from scientific code. Among the four models, Gemini achieved the highest accuracy and moderate agreement with human annotations. LLaMa has the lowest accuracy and agreement. Common misclassifications included acronyms, plural nouns, and dual-function words, highlighting challenges with contextual and domain-specific details.

\subsection*{\RQB}
\begin{table}
\centering
\caption{Number of times each LLM retained the original method name.}
\vspace{-3mm}
\label{Table:preserve-name}
\resizebox{\columnwidth}{!}{%
\begin{tabular}{@{}lrr@{}}
\toprule
\multicolumn{1}{c}{\multirow{2}{*}{\textbf{Model}}} & \multicolumn{2}{c}{\textbf{Preserved Original Name}}                         \\ \cmidrule(l){2-3} 
\multicolumn{1}{c}{}                                & \multicolumn{1}{c}{\textbf{Count}} & \multicolumn{1}{c}{\textbf{Percentage}} \\ \midrule
Gemini 2.0 Flash   & 324 & 65.3\% \\
Qwen 2.5 Coder 32B & 160 & 32.3\% \\
DeepSeek-R1 70B    & 136 & 27.4\% \\
LLaMa 3.3 70B      & 35  & 7.1\%  \\ \bottomrule
\end{tabular}%
}
\end{table}

\begin{table}
\centering
\caption{Length of method names generated by the LLMs.\\
The values within parentheses represent the growth relative to the original method name.}
\vspace{-3mm}
\label{Table:word-char-count}
\resizebox{\columnwidth}{!}{%
\begin{tabular}{@{}lrrrr@{}}
\toprule
\multicolumn{1}{c}{\multirow{2}{*}{\textbf{Model}}} &
  \multicolumn{2}{c}{\textbf{\begin{tabular}[c]{@{}c@{}}Average\\ Words Per Method Name\end{tabular}}} &
  \multicolumn{2}{c}{\textbf{\begin{tabular}[c]{@{}c@{}}Average\\ Characters Per Method Name\end{tabular}}} \\ \cmidrule(l){2-5} 
\multicolumn{1}{c}{} &
  \multicolumn{1}{c}{\textbf{\begin{tabular}[c]{@{}c@{}}Original\\ Method\end{tabular}}} &
  \multicolumn{1}{c}{\textbf{\begin{tabular}[c]{@{}c@{}}Corrected\\ Method\end{tabular}}} &
  \multicolumn{1}{c}{\textbf{\begin{tabular}[c]{@{}c@{}}Original\\ Method\end{tabular}}} &
  \multicolumn{1}{c}{\textbf{\begin{tabular}[c]{@{}c@{}}Corrected\\ Method\end{tabular}}} \\ \midrule
Gemini 2.0 Flash   & 1.78 & 2.77 (+55.62\%) & 10.86 & 17.97 (+65.47\%) \\
DeepSeek-R1 70B    & 1.94 & 2.96 (+52.58\%) & 11.29 & 19.96 (+76.79\%) \\
Qwen 2.5 Coder 32B & 1.87 & 2.99 (+59.89\%) & 10.88 & 19.77 (+81.71\%) \\
LLaMa 3.3 70B      & 1.98 & 3.29 (+66.16\%) & 11.71 & 22.62 (+93.17\%) \\ \bottomrule
\end{tabular}%
}
\end{table}
While the prior RQ focused on the grammar patterns generated by LLMs for the original practitioner-crafted method name, this RQ focuses on the extent to which LLMs evaluate the quality of the original name and suggest a correction. 

First, we analyze the agreement between the four LLMs by comparing the corrected names provided by each LLM. We observe that only 78 out of 496 corrected names (about 15.7\%) are identical across all four LLMs. Further, the level of agreement among the four LLMs is measured by a Fleiss' Kappa value of 0.389, indicating a fair level of consensus. 

Next, we examined the extent to which each LLM agreed with the original method name (i.e., the suggested corrected name was the same as the original name). As shown in Table \ref{Table:preserve-name}, Gemini showed the strongest preference for keeping original names unchanged, preserving about two-thirds (65.3\%) of the names. In contrast, LLaMA was the most aggressive in proposing alternative names, with approximately 93\% of its suggestions being different from the original.

Moving on, we next examine the structural characteristics of the suggested method names. We perform this analysis exclusively on the method names that were changed by each LLM. Table \ref{Table:word-char-count} shows the findings of this analysis. Since the number of method names corrected varied across each LLM, the values in the column labeled \texttt{Original Method} represent the average counts for only the original method names specific to each LLM. We observe that all four models tend to lengthen the method name. LLaMA produced the longest method names, with approximately 66\% more words and 93\% more characters than the original names on average. In contrast, Gemini shows the least growth of the corrected method name. Deepseek and Qwen contribute method names that are typically around 20 characters long and comprise approximately three words for their suggested renames.

Our next set of analyses explored the semantic characteristics of the names. We first start by investigating the common terms that each LLM adds to and removes from the original name when it suggests an alternate name. Table \ref{Table:added-removed} shows the top two frequently added and removed terms for each LLM\footnote{Due to space constraints, we limited the table to two terms for each action.}. Two observations we notice in all four LLMs are as follows: (1) the term ``calculate'' dominates all four LLMs as the most added term, highlighting the unique characteristic of scientific code, and (2) the expansion of the acronym ``mse''. Another interesting observation is the replacement of the term ``get'' with a synonym like ``extract'' and ``retrieve'' by most LLMs.

Examining the part-of-speech tags generated by the LLMs for the corrected names, we found that the majority of grammar patterns produced by all four LLMs begin with a verb. Specifically, 97.98\% of the grammar patterns from LLaMA, 96.37\% from Qwen, 95.56\% from DeepSeek, and 85.69\% from Gemini start with a verb. In contrast, approximately 55\% of the patterns in the original dataset begin with a verb.

\begin{table}
\centering
\caption{The two terms most frequently added and removed by each LLM in the corrected method name.}
\vspace{-3mm}
\label{Table:added-removed}
\resizebox{\columnwidth}{!}{%
\begin{tabular}{@{}llrl@{}}
\toprule
\multicolumn{1}{c}{\textbf{Action}} & \multicolumn{1}{c}{\textbf{Term}} & \multicolumn{1}{c}{\textbf{Count}} & \multicolumn{1}{c}{\textbf{Example}} \\ \midrule
\multicolumn{4}{c}{Model: \textit{Gemini 2.0 Flash}}                                             \\
Added   & calculate & 51 & variance → calculate\_variance                               \\
Added   & image     & 12 & pdiguide\_imgRead → read\_pdiguide\_image                    \\
Removed & mse       & 8  & MSE → calculate\_mean\_squared\_error                        \\
Removed & im        & 7  & im\_convert → convert\_image                                 \\ \midrule
\multicolumn{4}{c}{Model: \textit{Qwen 2.5 Coder 32B}}                                           \\
Added   & calculate & 86 & square → calculate\_square                                   \\
Added   & get       & 21 & author\_url → get\_author\_url                               \\
Removed & mse       & 15 & MSE → calculate\_mean\_squared\_error                        \\
Removed & get       & 12 & get\_params → initialize\_parameters                         \\ \midrule
\multicolumn{4}{c}{Model: \textit{DeepSeek-R1 70B}}                                              \\
Added   & calculate & 55 & pcr\_noise → calculate\_pcr\_noise                           \\
Added   & compute   & 20 & \_hash → compute\_hash                 \\
Removed & get       & 21 & get\_dataset → retrieve\_dataset           \\
Removed & mse       & 15 & MSE → mean\_squared\_error                                   \\ \midrule
\multicolumn{4}{c}{Model: \textit{LLaMa 3.3 70B}}                                                \\
Added   & calculate & 91 & square → calculate\_square                                   \\
Added   & perform   & 27 & preprocess → perform\_preprocessing  \\
Removed & get       & 37 & get\_year → extract\_year\_from\_name \\
Removed & mse       & 15 & MSE → calculate\_mean\_squared\_error                        \\ \bottomrule
\end{tabular}%
}
\end{table}

Further, as part of the corrected name grammar pattern analysis, we observed a discrepancy between the number of words in a method name and the number of part-of-speech tags in the grammar pattern. From Table \ref{Table:tag-word-count}, we observe that Gemini shows the highest consistency in matching part-of-speech tags to words, with a rate of 95.77\%. In contrast, LLaMA has the lowest matching rate (69.15\%) and generates fewer part-of-speech tags than there are words in the method name. For instance, LLaMA suggests renaming ``pcr\_noise'' to ``apply\_pcr\_noise\_model,'' which contains four words, but it generates only two tags (V,NM) for the corrected name. On the other hand, DeepSeek tends to assign more part-of-speech tags than words. For example, while the model suggests renaming ``update'' to ``train\_model,'' the new name is matched with a grammar pattern of three tags (V,N,N) rather than two.

\begin{table}
\centering
\caption{Consistency of word counts and part-of-speech tags from LLMs for the corrected method name.}
\vspace{-3mm}
\label{Table:tag-word-count}
\resizebox{\columnwidth}{!}{%
\begin{tabular}{@{}lrrr@{}}
\toprule
\multicolumn{1}{c}{\textbf{Model}} &
  \multicolumn{1}{c}{\textbf{\begin{tabular}[c]{@{}c@{}}Equal \\ Tags and Words\end{tabular}}} &
  \multicolumn{1}{c}{\textbf{\begin{tabular}[c]{@{}c@{}}More \\ Tags than Words\end{tabular}}} &
  \multicolumn{1}{c}{\textbf{\begin{tabular}[c]{@{}c@{}}Fewer \\ Tags than Words\end{tabular}}} \\ \midrule
Gemini 2.0 Flash   & 475 (95.77\%) & 19 (3.83\%) & 2 (0.40\%)    \\
DeepSeek-R1 70B    & 384 (77.42\%) & 30 (6.05\%) & 82 (16.53\%)  \\
Qwen 2.5 Coder 32B & 442 (89.11\%) & 22 (4.44\%) & 32 (6.45\%)   \\
LLaMa 3.3 70B      & 343 (69.15\%) & 20 (4.03\%) & 133 (26.81\%) \\ \bottomrule
\end{tabular}%
}
\end{table}

\begin{table}
\centering
\caption{Count of abbreviations and acronyms expanded by each LLM.}
\vspace{-3mm}
\label{Table:abbrev}
\resizebox{\columnwidth}{!}{%
\begin{tabular}{@{}lrr@{}}
\toprule
\multicolumn{1}{c}{\multirow{2}{*}{\textbf{Model}}} & \multicolumn{2}{c}{\textbf{\begin{tabular}[c]{@{}c@{}}Abbreviations \& \\ Acronyms\end{tabular}}} \\ \cmidrule(l){2-3} 
\multicolumn{1}{c}{}                                & \multicolumn{1}{c}{\textbf{Not Expanded}}         & \multicolumn{1}{c}{\textbf{Expanded}}         \\ \midrule
Gemini 2.0 Flash   & 88 (62.86\%) & 52 (37.14\%)  \\
Qwen 2.5 Coder 32B & 58 (41.43\%) & 82 (58.57\%)  \\
DeepSeek-R1 70B    & 42 (30\%)    & 98 (70\%)     \\
LLaMa 3.3 70B      & 38 (27.14\%) & 102 (72.86\%) \\ \bottomrule
\end{tabular}%
}
\end{table}

Our final analysis examines how the LLMs handle method names containing abbreviations/acronyms. The original dataset of 496 method names contains 140 instances of such terms. 

First, we examine how many of these 140 methods were not corrected (i.e., the LLM did not propose an alternative). We observe that Gemini preserved the original name in 53 instances (approximately 37.9\%). In comparison, Qwen, DeepSeek, and LLaMA preserved the original name 8 times (5.7\%), 6 times (4.3\%), and 1 time (0.7\%), respectively.

Next, from Table \ref{Table:abbrev}, we note that Gemini is more conservative and expands only about 37.14\% of the abbreviations and acronyms. In contrast, LLaMA is more aggressive, expanding 72.86\%. Some observations we notice are that certain computing abbreviations, such as `CSV', `JSON', `ASCII', and `PNG' are not expanded by any of the LLMs. Similarly, we find that some domain-specific abbreviations, like `PCR' (Polymerase Chain Reaction) `FES' (Filter Encoding Standard), and `WISDM' (Wireless Sensor Data Mining), are not expanded by all four LLMs. In contrast, abbreviations like `MSE' (Mean Squared Error) and acronyms like `ACC' (Accuracy) are expanded by most of the LLMs.

\vspace{1mm}
\noindent\textbf{Summary.}
RQ2 assesses LLMs' ability to suggest method name corrections that adhere to best practices. Gemini retained 65.3\% of original names, while LLaMA changed 93\%. Corrected names were longer and often started with verbs, with common additions included terms like ``calculate.'' LLaMA is the most aggressive among the LLMs in expanding abbreviations and acronyms. However, computing and domain-specific abbreviations, such as ``CSV'' and ``PCR'' were often not expanded in the suggested name.    

\section{Discussion}
\label{Section:discussion}
As an exploratory study, our findings show that while LLMs can provide valuable guidance for improving scientific code readability, their suggestions require careful human evaluation, particularly in domain-specific contexts.

A clear observation from our RQ results is the unique behavior of the individual LLMs in our evaluation. Specifically, there are significant differences in how Gemini and LLaMA generate grammatical patterns and assess the quality of names. This shows that different architectural approaches and training methodologies can lead to fundamentally different code analysis philosophies. Moreover, the challenges with the LLMs in handling domain-specific terminology show that the assumptions built into current LLM architectures regarding code quality may not align with the needs of scientific computing. This highlights the need for explicit domain-aware training approaches instead of relying on general-purpose language models for specialized code analysis tasks.

The results we achieved are a result of the prompt we created, which was developed through multiple iterations and an in-depth understanding of software engineering principles and the behavior patterns of LLMs. It is important to acknowledge that research scientists who lack a background in software engineering are unlikely to replicate our level of success. While they might apply basic prompts like ``improve this method name'' or ``make this code better,'' these simplistic approaches may yield poorer results than our findings. This potential gap between our controlled experimental conditions and real-world usage scenarios is a concern. Hence, there is a need to make LLMs more accessible and effective for users across varying levels of technical expertise.

\noindent\textbf{Key Implications For \textit{SE Research Community}.} Our work advances the field of AI for SE by providing an initial evaluation of LLMs for improving program comprehension in scientific code. Our early results highlight several opportunities for further research in this area, such as domain-specific fine-tuning approaches for LLMs tailored to specific scientific disciplines and conducting more human-centric studies to understand how research scientists utilize LLMs.

\noindent\textbf{Key Implications For \textit{Research Scientists}.} While non-computer science researchers may be inclined to rely entirely on LLMs to write code that supports their research, our findings provide guidance on when to trust LLM suggestions and when human judgment is essential. Thus, helping them make more informed decisions regarding adopting AI coding assistants. Scientists can utilize these insights to better evaluate code improvements generated by LLMs, especially as domain-specific terminology often needs human review for accuracy

\noindent\textbf{Key Implications For \textit{Educators}.} When teaching programming concepts to students from non-traditional computer science backgrounds, educators should emphasize both the advantages and limitations of using AI tools for software engineering. Helping students understand when and why LLMs may fail in code analysis tasks is essential for developing their critical evaluation skills in AI-assisted development.

\section{Threats To Validity}
\label{Section:threats}
The prompt's quality can significantly impact LLM output, potentially affecting our findings.  
Additionally, since we applied the same prompt across all four LLMs, there is a risk that this prompt may not have been ideal for every model.

As LLMs continue to evolve, the results observed in our study may not reflect the capabilities of future models. Our findings should therefore be considered as a snapshot of current LLM capabilities rather than an assessment of their potential for code analysis tasks. Additionally, due to the non-deterministic nature of LLM outputs, there is a threat to reproducing our results. Furthermore, the selected LLMs may not represent all available models, which could also limit the applicability of our results. However, since this is an exploratory study, our goal is to identify preliminary trends and patterns that can guide future research.

Our dataset is limited to a subset of methods found in Python-based Jupyter Notebooks, which restricts the generalizability of our findings. We also acknowledge that method names were analyzed in isolation, without additional context such as class names, file names, or usage patterns. While this simplifies the analysis and reduces prompt complexity, it may limit the relevance or precision of the suggested names, especially in cases where meaning is derived from surrounding structures. Finally, we did not include human evaluation of method name corrections generated by LLMs, limiting our assessment of semantic appropriateness. Expert feedback could enhance the evaluation's real-world applicability.

\section{Conclusion \& Future Work}
\label{Section:conclusion}
This exploratory study examines the effectiveness of LLMs in analyzing and improving method names in scientific code. Our evaluation of four popular LLMs on 496 method names shows notable differences in LLM performance for grammatical pattern recognition and method name improvements. Gemini achieved the highest accuracy in recognizing grammatical patterns, but a moderate agreement with human annotations. For method name corrections, Gemini was conservative and preserved most of the original names, while LLaMA made more aggressive changes. All models showed a bias toward names that begin with a verb, aligning with software engineering practices. As an initial investigation into the use of LLMs for improving program comprehension in scientific code, our findings establish a foundation for future research in this area. Our future work will include other identifier types, such as variables, parameters, and classes, for a comprehensive understanding of identifier quality in scientific code.

\section{Acknowledgments}
\label{Section:acknowledgments}
The technical support and advanced computing resources from University of Hawaii Information Technology Services – Research Cyberinfrastructure, funded in part by the National Science Foundation CC* awards \# 2201428 and \# 2232862 are gratefully acknowledged. Additionally, this work was partially supported by the Undergraduate Research Opportunities Program in the Office of the Vice Provost for Research and Scholarship at the University of Hawaiʻi at Mānoa.

\bibliographystyle{ieeetr}
\bibliography{main}
\end{document}